\documentclass[aps,prd,preprintnumbers,showpacs,superscriptaddress,nofootinbib,amsmath,amssymb,floats,floatfix,showkeys,notitlepage,longbibliography]{revtex4-1}
\usepackage[table]{xcolor} % For table row colors
\usepackage{array} % For custom column alignments
\renewcommand{\arraystretch}{1.5} % Adjust row height for better readability
\definecolor{lightblue}{rgb}{0.62, 0.83, 0.96} % Light blue color
\definecolor{lightgray}{rgb}{0.93, 0.97, 0.99} % Light gray color
\usepackage{comment}
\usepackage{graphicx}
\usepackage{subfigure}
\usepackage{palatino}
\usepackage[commandnameprefix=always]{changes}
\usepackage{hyperref}
\hypersetup{colorlinks=true,linkcolor=red,urlcolor=red,citecolor=red}
\usepackage[toc,page]{appendix}
\usepackage[normalem]{ulem}
\usepackage{lipsum}
\usepackage{graphicx}
\usepackage{subfigure}
\usepackage{palatino}
\usepackage{float}
\usepackage{sans}
\usepackage{adjustbox}
\usepackage{latexsym}
\usepackage{amsmath}
\usepackage{amssymb}
\usepackage{amsfonts}
\usepackage{dcolumn}
\usepackage{bm}
\usepackage{tikz}
\usepackage{bigints}
\usepackage{array,tabularx,multirow,booktabs}
\usepackage[tracking=true]{microtype}
\SetTracking{}{500}
\SetTracking{encoding={*}, shape=sc}{40}
\allowdisplaybreaks
\usepackage{adjustbox}
\usepackage{latexsym}
\usepackage{amsmath}
\usepackage{amssymb}
\usepackage{amsfonts}
\usepackage{dcolumn}
\usepackage{bm}
\usepackage{tikz}
\usepackage{bigints}
\usepackage{array,tabularx,multirow,booktabs}
\usepackage[tracking=true]{microtype}
\usepackage{color}
\usepackage[utf8]{inputenc}
\usepackage{amsmath}
\usepackage{array}
\usepackage{booktabs}
\usepackage{multirow}
\usepackage{caption}
\usepackage{xcolor}
\usepackage{colortbl}
\usepackage{geometry}

\definecolor{headerblue}{HTML}{003366}
\definecolor{rowgray}{HTML}{F8F9FA}
\allowdisplaybreaks

\begin{document}

\title{Holographic CFT Phase Transitions and Criticality for
Einstein-Maxwell-Power-Yang-Mills AdS Black Holes\\\;\\\;\\}

\author{Mohammad Reza Alipour}
\email{mr.alipour@stu.umz.ac.ir}
\affiliation{School of Physics, Damghan University, P. O. Box 3671641167, Damghan, Iran}

\author{Mohammad Ali S. Afshar}
\email{m.a.s.afshar@gmail.com}
\affiliation{School of Physics, Damghan University, P. O. Box 3671641167, Damghan, Iran}

\author{Saeed Noori Gashti}
\email{sn.gashti@du.ac.ir; saeed.noorigashti70@gmail.com}
\affiliation{School of Physics, Damghan University, P. O. Box 3671641167, Damghan, Iran}

\author{Behnam Pourhassan}
\email{b.pourhassan@du.ac.ir}
\affiliation{School of Physics, Damghan University, P. O. Box 3671641167, Damghan, Iran}

\affiliation{Center for Theoretical Physics, Khazar University, 41 Mehseti Street, Baku, AZ1096, Azerbaijan}

\begin{abstract}
We present a comprehensive study of the thermodynamic phase structure for Anti-de Sitter black holes in Einstein-Maxwell-power-Yang-Mills gravity, reformulated through holographic duality as an ensemble problem in the dual conformal field theory (CFT). By deriving an extended first law where the central charge \(C\) is a thermodynamic variable, we systematically explore both canonical and mixed ensembles. In the canonical ensemble with fixed charges, we identify a van der Waals-like phase transition between small and large black holes, marked by a characteristic swallowtail structure and coexistence curves with a negative slope. In contrast, within the mixed ensemble of fixed electric potential, the system exhibits a Hawking-Page transition between confined and deconfined phases of the boundary CFT. Our key finding is the suppressive role of the non-Abelian Yang-Mills charge \(\tilde{q}\): increasing \(\tilde{q}\) lowers both the minimum and the Hawking-Page transition temperatures, significantly narrowing the stability window of the confined phase. These results, supported by detailed numerical analysis, reveal a rich, ensemble-dependent phase landscape and establish the non-linear Yang-Mills sector as a critical controller of confinement physics in strongly coupled holographic systems.
\begin{center}
\end{center}
\date{\today}
\begin{center}
\end{center}
\end{abstract}
\maketitle
\tableofcontents
\section{Introduction} \label{isec1}
We begin with a foundational question: What is the significance of studying the thermodynamics of conformal field theories (CFTs) in the context of black holes? This query is deeply rooted in the development of quantum gravity, particularly through the holographic principle \cite{1} and the anti–de Sitter/conformal field theory (AdS/CFT) correspondence \cite{2}. Black holes are not merely simple gravitational wells; they are complex quantum systems endowed with a vast number of microscopic degrees of freedom, as quantified by the Bekenstein-Hawking entropy,  
$ S_{\text{BH}} = k_B A/4\ell_P^2$ \cite{3,4}. Consequently, any prospective theory of quantum gravity aspiring to unify fundamental physics must be able to incorporate a precise and logical outcomes of the statistical mechanical concepts resulting from this entropy by identifying and enumerating these underlying microstates \cite{5,5',5''}. Without such a microscopic accounting, black hole entropy remains a purely phenomenological construct.\\
In this regard, and to prevent such a defect, there was a golden key in past studies that showed, explicit quantum mechanical frameworks, such as two-dimensional CFTs, can elucidate the microscopic origin of $ S_{\text{BH}} $ and reproduce its thermal value \cite{6,7}. This positions the thermodynamic analysis of dual 2D CFTs as a uniquely powerful probe into black hole physics. In holographic settings, broad classes of rotating and charged black holes admit dual 2D CFT descriptions \cite{8,8',9,9',10,10',10'',10'''}, where macroscopic thermodynamic relations (e.g., mass, angular velocity, electric potential) map precisely to CFT quantities (e.g., left/right temperatures $T_L, T_R$ and central charges $c_L, c_R$) \cite{11,12}. This correspondence sharpens the microscopic interpretation of $ S_{\text{BH}} $ through exact Cardy formula state counting \cite{13} and clarifies how horizon mechanics emerges from universal CFT properties.
CFT thermodynamics offers a window into black hole microstate dynamics. Phenomena such as the Hawking-Page transition \cite{14}—interpreted in the dual CFT as a confinement-deconfinement phase transition \cite{15}—illustrate how black hole formation and evaporation manifest microscopically. Moreover, thermal correlators and quasinormal mode spectra in CFTs encode real-time dynamics and black hole relaxation timescales \cite{16}, bridging macroscopic stability with microscopic thermalization. The laws of thermodynamics at inner and outer horizons collectively encode the dual CFT’s global data (e.g., $T_{Left}$, $T_{Right}$, $c_L$, $c_R$) without requiring detailed microphysical input \cite{12,17}. Crucially, deformations probing distinct conserved charges (e.g., electric, magnetic, NUT charges) correspond to distinct dual CFT sectors, enabling consistent holographic images of the black hole. This horizon-based thermodynamic encoding provides robust CFT signatures and a direct pathway to extract CFT parameters from gravitational data \cite{8,8',9,9',10,10',10'',10''',11}.  Collectively, these insights establish that CFT thermodynamics transcends mathematical curiosity: it constitutes the dual language through which black hole quantum physics is articulated in holographic quantum gravity \cite{2}. By studying it, physicists access non-gravitational descriptions of black hole microstates, phase structure, and quantum information dynamics (elements indispensable to unifying quantum gravity). Ultimately, CFT thermodynamics furnishes a rigorous testing ground where gravitational entropy, information flow, and emergent spacetime geometry are encoded in the renormalization group flow and entanglement structure of a unitary quantum system \cite{18,18',18'',18''',18''''}. Nevertheless, significant challenges persist. Unlike gauge fields (e.g., electromagnetism), which remain well-defined under holographic mapping, gravity requires a precise dictionary to translate bulk quantities to boundary CFT observables \cite{15,19}. Constructing such dictionaries, especially for nonlinear or modified gravity theories, is often nontrivial, hindering the harmonious development of CFT thermodynamic analyses \cite{20,21,22,23,24} compared to bulk studies \cite{25,25a,25b,26,26a,26b,27,27a,27b,28,28a,28b,29,29a,29b,29c,29d}.\\    
In this paper, we address these challenges by examining an asymptotically AdS black hole solution in Einstein-Maxwell-power-Yang-Mills (EMPYM) gravity \cite{30,30a}.
EMPYM gravity extends general relativity by coupling spacetime geometry to two distinct gauge fields:\\  
(i) an Abelian $U(1)$ Maxwell field\\
(ii) a non-Abelian $SU(N)$ Yang-Mills field with a power-law-modified kinetic term $\mathcal{L}_{\text{YM}} \propto (\text{Tr} F_{\mu\nu}F^{\mu\nu})^\gamma$ ($\gamma \neq 1$) \cite{31}. 
This framework provides a rich laboratory for probing nonlinear gauge interactions in strong-gravity regimes. The power-law deformation opens up the field and possibility of extensive research on reexamining singularities, black hole thermodynamic behaviors, and energy conditions and cosmic censorship. While phenomenological viability requires further study, EMPYM gravity serves as a critical testbed for quantum gravity conjectures where gauge and gravitational degrees of freedom become non-perturbatively intertwined \cite{32,32a,32b,32c}. Leveraging a derived CFT dictionary, in this work, we perform a first-principles analysis of this black hole’s CFT thermodynamics, bridging gravitational dynamics with its dual quantum statistical mechanics.\\\\
This research presents a coherent and comprehensive analysis of the thermodynamic properties and phase structure of Anti-de Sitter (AdS) black holes within Einstein-Maxwell-power-Yang-Mills gravity, employing the principles of holographic duality to interpret the observed phenomena in the language of the dual boundary conformal field theory (CFT). The article is structured to guide the reader systematically from the foundational gravitational description to a detailed analysis of phase transitions in the dual field theory.\\
The work commences with a meticulous review of black hole solutions in \(N\)-dimensional spacetime. In this foundational section, the fundamental action of the theory, the corresponding field equations, and ultimately the explicit metric for static, spherically symmetric black holes are presented. Key parameters of the model, such as mass (\(M\)), electromagnetic charge (\(Q\)), non-Abelian Yang-Mills charge (\(q\)), and the power parameter (\(\gamma\)) controlling the nonlinearity of the Yang-Mills sector, are defined within this framework. From this metric, essential thermodynamic quantities of the black hole in the bulk spacetime, including Hawking temperature (\(T\)), Bekenstein-Hawking entropy (\(S\)), and electric potential (\(\phi\)), are computed. This stage solidifies the technical foundation and precisely defines the gravitational object under study.
Subsequently, by crossing the conceptual bridge of holography, the thermodynamics of this black hole are translated into the language of its dual boundary field theory. This translation is accomplished using the standard holographic dictionary and introducing a conformal factor (\(\omega\)) for generalized boundary scaling. In this process, bulk quantities are mapped to their corresponding dimensionless boundary CFT variables (\(\tilde{E}, \tilde{T}, \tilde{Q}, \tilde{q}, \tilde{\phi}, \tilde{\psi}\)). The core of this section is the derivation of the extended first law of thermodynamics for the boundary field theory, expressed as
$
d\tilde{E} = \tilde{T} dS + \tilde{\phi} d\tilde{Q} + \tilde{\psi} d\tilde{q} + \mu dC - p d\mathcal{V}.\\
$
In this relation, the new and crucial variable central charge (\(C\)), characteristic of the boundary field theory, emerges as an independent thermodynamic quantity, introducing its conjugate chemical potential (\(\mu\)) and the boundary pressure (\(p\)) into the first law. This equation forms the fundamental identity for all subsequent analyses.
Following the establishment of this framework, the article proceeds to analyze local stability and identify critical behavior. By examining the boundary temperature (\(\tilde{T}\)) as a function of entropy (\(S\)), the mathematical conditions for the emergence of critical points (\(\partial\tilde{T}/\partial S = 0\) and \(\partial^2\tilde{T}/\partial S^2 = 0\)) are applied. Solving these conditions leads to obtaining explicit relations for the critical entropy (\(S_c\)), critical Yang-Mills charge (\(\tilde{q}_c\)), and critical temperature (\(\tilde{T}_c\)). The parametric domain for the existence of these critical points, particularly its dependence on the nonlinear parameter \(\gamma\), is carefully investigated. Furthermore, calculating the heat capacity (\(\tilde{\mathcal{C}}\)) and examining its sign delineates the regions of local stability and instability of the system, revealing the parameter range where the system is susceptible to phase transitions.\\
The article then advances to its primary exploration, conducting a global and detailed analysis of the phase structure within different thermodynamic ensembles. This section demonstrates the richness of the system's phase behavior. In the canonical ensemble, where the electric and Yang-Mills charges (\(\tilde{Q}, \tilde{q}\)) are held fixed, the relevant potential is the Helmholtz free energy (\(\tilde{F} = \tilde{E} - \tilde{T}S\)). Studying this potential reveals that for a certain parameter range, the system exhibits behavior remarkably similar to a van der Waals fluid. The free energy plots unveil the characteristic swallowtail structure, indicative of a first-order phase transition between small black hole (SBH) and large black hole (LBH) phases. This structure disappears at a critical point where the transition becomes second-order. Mapping the coexistence curves in the \(\tilde{Q}\)-\(\tilde{T}\) and \(\tilde{q}\)-\(\tilde{T}\) planes is also performed, with attention drawn to their distinctive negative slope.\\
In contrast, within a mixed or semi-grand canonical ensemble, where the electric potential (\(\tilde{\phi}\)) is fixed instead of its conjugate charge, a completely different phase phenomenon occurs. The thermodynamic potential for this ensemble is \(\tilde{W} = \tilde{E} - \tilde{T}S - \tilde{\phi}\tilde{Q}\). Investigation of this function reveals the presence of a Hawking-Page-type phase transition between a confined and a deconfined phase. For values of \(\tilde{\phi}\) below a maximum, the free energy exhibits a cusp at a minimum temperature (\(\tilde{T}_0\)) and changes sign from positive to negative at a higher temperature (\(\tilde{T}_{HP}\)). This sign change signifies a first-order transition from the stable confined phase (which can correspond to the AdS thermal radiation background) to the stable deconfined phase (corresponding to the large black hole). A key quantitative finding of this article emerges here: the computed numerical data, presented in detailed tables, clearly demonstrate that increasing the Yang-Mills charge (\(\tilde{q}\)) weakens and suppresses this Hawking-Page transition. Holding other parameters constant, increasing \(\tilde{q}\) causes a decrease in both temperatures \(\tilde{T}_0\) and \(\tilde{T}_{HP}\) and systematically narrows the temperature window (\(\Delta\tilde{T}\)) associated with the confined phase.\\
Finally, the article concludes with a comprehensive synthesis. The principal findings, namely, the discovery of two distinct, ensemble-dependent phase landscapes (van der Waals-like and Hawking-Page-like), are reviewed and juxtaposed. Emphasis is placed on the central physical insight of the work: the nonlinear Yang-Mills charge (\(\tilde{q}\)) acts as a potent suppressor of the Hawking-Page (confinement) transition. This result carries significant implications for understanding strongly coupled gauge theories with non-Abelian degrees of freedom. The article compares the influence of various parameters such as \(\tilde{Q}\), \(\tilde{q}\), and \(C\) on the phase diagrams across different ensembles and, ultimately, outlines promising directions for future research. These include investigating the third logical ensemble (with fixed \(\tilde{Q}, \tilde{\psi}\)), studying dynamical stability and tunneling rates, searching for triple points in the phase diagram, and generalizing this analysis to other spacetime dimensions or different horizon topologies.
\section{Review of Black Hole Solutions in $N$-Dimensional Einstein-Maxwell-Power-Yang-Mills Gravity}
In this section, we present an overview of black hole solutions in $N$-dimensional Einstein-Maxwell-Power-Yang-Mills (EMPYM) gravity with a cosmological constant $\Lambda$. The action integral for this model is given by,
\begin{equation}\label{eq1}
\mathcal{S} = \frac{1}{2} \int dx^N \sqrt{-g} \left[ R - \frac{(N-1)(N-2)}{3} \Lambda - F_{\mu\nu}F^{\mu\nu} - \left( \text{Tr}(F_{\mu\nu}^{(a)} F^{(a)\mu\nu}) \right)^\gamma \right],
\end{equation}
where $R$ is the Ricci scalar, $\gamma$ is a positive real parameter controlling the nonlinearity of the Yang-Mills term, and the trace operator is defined as,
$\text{Tr}(.) = \sum_{a=1}^{\frac{(N-1)(N-2)}{2}} (.)$.
The field strength tensors for the Yang-Mills and Maxwell fields are respectively defined as,
\begin{equation}\label{eq2}
F_{\mu\nu}^{(a)} = \partial_\mu A_\nu^{(a)} - \partial_\nu A_\mu^{(a)} + \frac{1}{2\sigma} C_{(b)(c)}^{(a)} A_\mu^{(b)} A_\nu^{(c)},
\end{equation}
\begin{equation}\label{eq3}
F_{\mu\nu} = \partial_\mu A_\nu - \partial_\nu A_\mu,
\end{equation}
where $C_{(b)(c)}^{(a)}$ are the structure constants of a Lie group $G$ with dimension $\frac{(N-1)(N-2)}{2}$, $\sigma$ is the Yang--Mills coupling constant, $A_\mu^{(a)}$ are the gauge potentials associated with the $SO(N-1)$ group, and $A_\mu$ is the standard Maxwell potential.
The spacetime geometry is assumed to be spherically symmetric, with the line element given by,
\begin{equation}\label{eq4}
ds^2 = -f(r)\,dt^2 + \frac{dr^2}{f(r)} + r^2 d\Omega_n^2,
\end{equation}
where $d\Omega_n^2$ denotes the metric of the unit $n$-sphere,  $N = n + 2 \geq 4$, and $\gamma \neq \frac{n+1}{4}$.  The metric function $f(r)$ for the EMPYM black hole with negative cosmological constant $\Lambda=-\frac{n(n+1)}{2\ell^2}$ is given by,
\begin{equation}\label{eq5}
f(r) = 1 - \frac{2GM}{r^{n-1}} + \frac{n(n+1)}{6} \frac{r^2}{\ell^2} + \frac{2(n-1)Q^2}{n r^{2n-2}} - \frac{\mathcal{Q}}{r^{4\gamma - 2}},
\end{equation}
with the Yang-Mills charge term defined as
\begin{equation}\label{eq6}
\mathcal{Q} = \frac{G[(n-1)n q^2]^{\gamma}}{n(4\gamma - n - 1)}.
\end{equation}
Here, $M$ represents the ADM mass of the black hole, $Q$ is the Maxwell field charge, and $q$ is the Yang--Mills charge parameter.
In this framework, the Hawking temperature, the black hole mass, and the entropy can be formulated in terms of the horizon radius $r_{h}$ within the extended phase space. Throughout our analysis, we restrict ourselves to the case $n=2$, which corresponds to a four-dimensional spacetime $N=4$.
\begin{equation}\label{eq7}
\begin{split}
T=\frac{f^{\prime}(r_h)}{4\pi}=-\frac{2^{\gamma -3} G q^{2 \gamma } r_h^{1-4 \gamma }}{\pi }+\frac{3 r_h}{4 \pi  \ell^2}-\frac{Q^2}{4 \pi  r_h^3}+\frac{1}{4 \pi  r_h}
\end{split}
\end{equation}

\begin{equation}\label{eq8}
\begin{split}
M=\frac{r_h^3}{2 G \ell^2}+\frac{Q^2}{2 G r_h}+\frac{r_h}{2 G}+\frac{2^{\gamma -2} q^{2 \gamma } r_h^{3-4 \gamma }}{4 \gamma -3}
\end{split}
\end{equation}

\begin{equation}\label{eq9}
\begin{split}
S=\frac{\pi  r_h^2}{G},    
\end{split}
\end{equation}
Also,the electric potential $\phi$, and the Yang-Mills potential  $\psi$,  can be expressed as,
\begin{equation}\label{eq10}
\begin{split}
 \phi =\frac{Q}{G r_h} \qquad \psi =\frac{2^{\gamma -1} \gamma  q^{2 \gamma -1} r_h^{3-4 \gamma }}{4 \gamma -3}
\end{split}
\end{equation}

In the framework of the extended phase space, the thermodynamic quantities satisfy the first law of black hole thermodynamics. By applying a scaling argument, the generalized Smarr relation for the EMPYM black hole can be consistently derived within this extended phase space,
\begin{equation}\label{eq11}
\begin{split}
dM=T dS  +\phi dQ  + \psi dq +V dP
\end{split}
\end{equation}
\begin{equation}\label{eq12}
\begin{split}
M=2 T S +\frac{(2 \gamma -1)  }{\gamma }q \psi+\phi Q  -2 P V
\end{split}
\end{equation}

\section{ Thermodynamics of the Boundary CFT in Einstein-Maxwell-Power-Yang-Mills-AdS Black Holes} \label{isec3}
We now aim to reformulate the bulk thermodynamic first law (2.22) in terms that explicitly involve the boundary central charge. To this end, we employ the holographic dual relation connecting the central charge $C$ , the AdS curvature scale , and Newton’s constant $G$ within Einstein gravity. In this framework, the central charge can be expressed as \cite{104,105,106,107,108,109}, 
  \begin{equation}\label{eq13}
\begin{split}
C=\frac{\Omega_2 \ell^2}{16\pi G} \quad \text{with} \quad \Omega_2=4\pi \quad \Rightarrow  \quad C=\frac{ \ell^2}{4 G}
\end{split}
\end{equation}
 with $\Omega_2=4\pi$ denoting the volume of the unit two-sphere. This correspondenceallows the extended first law to be rewritten in a boundary‑adapted form, where the central charge naturally appears as a fundamental thermodynamic variable of the dual CFT.

 In this section, we formulate the mass relation for the boundary CFT thermodynamics associated with AdS black holes in Einstein-Maxwell-power-Yang-Mills gravity. The induced metric on the boundary CFT is taken as \cite{104,105,106,107,108,109}: 
  \begin{equation}\label{eq14}
\begin{split}
dS^2=\omega^2(-dt^2+\ell^2 d\Omega_2^2)
\end{split}
\end{equation}
 where $\omega$ is a dimensionless conformal factor and $d\Omega_2^2$ denotes the line element of a unit two-sphere. While previous studies often fix $\omega=R/\ell$, with $R$ representing the curvature radius of the boundary, we retain $\omega$ as a free parameter to accommodate recent developments in holographic thermodynamics. This generalization enables the construction of a holographic first law that mirrors the extended first law of black hole thermodynamics, where the cosmological constant $\Lambda$ is treated as a dynamical variable and Newton’s constant $G_N$ remains fixed.
The spatial volume of the boundary sphere scales with , and for definiteness, we define it as \cite{104,105,106,107,108,109},
  \begin{equation}\label{eq15}
\begin{split}
\mathcal{V}=4 \pi  R^2
\end{split}
\end{equation}
Corresponding to this volume, a boundary pressure $p$ is introduced, contributing a work term $-pd \mathcal{V}$ in the first law. The holographic dictionary connecting bulk and boundary quantities is given by,
\begin{equation}\label{eq16}
\begin{split}
&E=M\frac{\ell }{R}=\frac{M}{\omega}, \quad \tilde{S}=S,  \quad \tilde{T}=T\frac{\ell }{R}=\frac{T}{\omega}, \quad\tilde{Q}=Q\frac{\ell }{\sqrt{G}}=2Q\sqrt{C}\\& \quad \tilde{\phi}=\frac{\phi }{R}, \quad \tilde{q}=q\frac{\ell }{\sqrt{G}}=2q\sqrt{C}, \quad 
\tilde{\psi}=\frac{\psi }{R}
\end{split}
\end{equation}
where $M,T,S,Q,\phi,q,\psi$ are the bulk mass, temperature, entropy, charge and electric potential,  charge and Yang-Mills potential, respectively, and their boundary counterparts are appropriately rescaled. Since Newton’s constant $G$ is treated as fixed in the framework of CFT thermodynamics, we shall set $G=1$ in the subsequent analysis.
Utilizing the extended thermodynamic relations and the volume definition above, the internal energy of the boundary CFT is expressed as,
\begin{equation}\label{eq17}
\begin{split}
E=\frac{2^{-\gamma -1} \pi ^{2 \gamma -\frac{3}{2}} C^{\frac{1}{2}-\gamma } S^{\frac{3}{2}-2 \gamma } \tilde{q}^{2 \gamma }}{(4 \gamma -3) R}+\frac{\sqrt{\pi } \tilde{Q}^2}{4 \sqrt{C} R \sqrt{S}}+\frac{S^{3/2}}{4 \pi ^{3/2} \sqrt{C} R}+\frac{\sqrt{C} \sqrt{S}}{ \sqrt{\pi } R}
\end{split}
\end{equation}
Given that $\omega=\frac{R}{\ell}$, we obtain,
 \begin{equation}\label{eq18}
\begin{split}
\frac{d\omega}{\omega}=\frac{dR}{R}-\frac{d\ell}{\ell}, \qquad  \frac{d\ell}{\ell}=\frac{1}{2}\frac{d \ell^2}{\ell^2}, \qquad \frac{dR}{R}=\frac{1}{2}\frac{d R^2}{R^2}
\end{split}
\end{equation}
By making use of equations \eqref{eq11}, \eqref{eq12}, \eqref{eq13}, \eqref{eq15}, \eqref{eq16}, and \eqref{eq18}, the first law of thermodynamics may be rewritten in the form:
\begin{equation}\label{eq19}
\begin{split}
d\bigg(\frac{M}{\omega}\bigg)=\frac{T}{\omega} dS+\frac{\phi}{R} d(Q \ell)+\frac{\psi}{R}d(q \ell)+\bigg[\frac{M}{\omega}-\frac{T}{\omega}-\frac{\phi}{R}(Q\ell)-\frac{3\gamma-1}{2\gamma}\frac{\psi}{R}(q \ell)\bigg]\frac{dC}{C}-\frac{M}{2\omega}\frac{d\mathcal{V}}{\mathcal{V}}
\end{split}
\end{equation}
Consequently, by applying equations \eqref{eq16} and \eqref{eq19}, the first law of thermodynamics in holographic space is expressed in the form:
 \begin{equation}\label{eq20}
\begin{split}
dE=\tilde{T} dS  +\tilde{\phi} d\tilde{Q}  + \tilde{\psi} d\tilde{q} +\mu dC-p d \mathcal{V}
\end{split}
\end{equation}
In this expression, $\mu$ corresponds to the chemical potential associated with the parameter $C$, while $p$ denotes the pressure in the field theory and $\mathcal{V}$ the spatial volume supporting the CFT, and we have,
\begin{equation}\label{eq21}
\begin{split}
\mu=\frac{1}{C}\bigg(E-\tilde{T} S  -\tilde{\phi} \tilde{Q}  - \frac{(3 \gamma -1)  }{2\gamma }\tilde{\psi} \tilde{q} \bigg),  \qquad   p=\frac{E}{2\mathcal{V}}  
\end{split}
\end{equation}
Furthermore, by employing equations \eqref{eq19} and \eqref{eq20}, the thermodynamic quantities
of the black hole in holographic space can be obtained,

\begin{equation}\label{eq22}
\begin{split}
\tilde{T}=&\bigg(\frac{\partial E}{\partial S}\bigg)_{\tilde{Q},\tilde{q},C,\mathcal{V}}=-\frac{2^{-\gamma -2} \pi ^{\frac{1}{2} (4 \gamma -1)-1} C^{\frac{1}{2}-\gamma } S^{\frac{1}{2} (1-4 \gamma )} \tilde{q}^{2 \gamma }}{R}-\frac{\sqrt{\pi } \tilde{Q}^2}{8 \sqrt{C} R S^{3/2}}+\frac{3 \sqrt{S}}{8 \pi ^{3/2} \sqrt{C} R}\\&+\frac{\sqrt{C}}{2 \sqrt{\pi } R \sqrt{S}},
\end{split}
\end{equation}

\begin{equation}\label{eq23}
\begin{split}
\tilde{\phi}=\bigg(\frac{\partial E}{\partial \tilde{Q}}\bigg)_{S,\tilde{q},C,\mathcal{V}}=\sqrt{\frac{\pi}{C S}}\frac{\tilde{Q}}{2R},
\end{split}
\end{equation}

\begin{equation}\label{eq24}
\begin{split}
\tilde{\psi}=\bigg(\frac{\partial E}{\partial \tilde{q}}\bigg)_{S,\tilde{Q},C,\mathcal{V}}=\frac{2^{-\gamma } \pi ^{2 \gamma -\frac{3}{2}} \gamma  S^{\frac{3}{2}-2 \gamma }}{(4 \gamma -3) R}C^{\frac{1}{2}-\gamma } \tilde{q}^{2 \gamma -1},
\end{split}
\end{equation}

\begin{equation}\label{eq25}
\begin{split}
\mu=&\bigg(\frac{\partial E}{\partial C}\bigg)_{S,\tilde{Q},\tilde{q},\mathcal{V}}=\frac{\sqrt{S}}{2 \sqrt{\pi C}  R}-\frac{S^{3/2}}{8 \pi ^{3/2} C^{3/2} R}+\frac{2^{-\gamma -1} \pi ^{2 \gamma -\frac{3}{2}} \left(\frac{1}{2}-\gamma \right) C^{-\gamma -\frac{1}{2}} S^{\frac{3}{2}-2 \gamma } \tilde{q}^{2 \gamma }}{(4 \gamma -3) R}\\&-\frac{\sqrt{\pi } \tilde{Q}^2}{8 C^{3/2} R \sqrt{S}},
\end{split}
\end{equation}
\begin{equation}\label{eq26}
\begin{split}
p=\bigg(\frac{\partial E}{\partial \mathcal{V}}\bigg)_{S,\tilde{Q},\tilde{q},C}=\frac{1}{8\pi R^2}\bigg(\frac{2^{-\gamma -1} \pi ^{2 \gamma -\frac{3}{2}} C^{\frac{1}{2}-\gamma } S^{\frac{3}{2}-2 \gamma } \tilde{q}^{2 \gamma }}{(4 \gamma -3) R}+\frac{\sqrt{\pi } \tilde{Q}^2}{4 \sqrt{C} R \sqrt{S}}+\frac{S^{3/2}}{4 \pi ^{3/2} \sqrt{C} R}+\frac{\sqrt{C} \sqrt{S}}{ \sqrt{\pi } R}\bigg)
\end{split}
\end{equation}
\section{ Critical points and  Heat capacity } \label{isec3}
The determination of the critical values of the thermodynamic variables is achieved by employing equation \eqref{eq20} in conjunction with the following set of relations,
\begin{equation}\label{eq27}
\begin{split}
\frac{\partial \tilde{T}}{\partial S}=0, \qquad \frac{\partial^2 \tilde{T}}{\partial S^2}=0
\end{split}
\end{equation}

\begin{equation}\label{eq28}
\begin{split}
S_c=-\frac{\pi ^2 (1-\gamma )}{2 \gamma }\bigg(\frac{4C}{6 \pi  (1-\gamma )}-\frac{4\gamma  C}{3 \pi  (1-\gamma )}-\sqrt{\frac{4 \gamma  \tilde{Q}_c^2}{\pi ^2 (1-\gamma )}+\left(\frac{4\gamma  C}{3 \pi  (1-\gamma )}-\frac{4C}{6 \pi  (1-\gamma )}\right)^2}\bigg)
\end{split}
\end{equation}
where,
\begin{equation}\label{eq29}
\begin{split}
&S_c=\frac{\pi  C}{3 \gamma }\bigg(2 \gamma -1+\sqrt{1+\frac{16 (\gamma -1) \gamma  \left(4 C^2-9 \tilde{Q}_c^2\right)}{C^2}}\bigg),\qquad \text{for} \quad \gamma  <1   \\
&S_c=\frac{2}{3 }\pi  C, \hspace{7.25cm} \text{for}\quad \gamma  =1\\
&S_c=\frac{\pi  C}{3 \gamma }\bigg(2 \gamma -1-\sqrt{1+\frac{16 (\gamma -1) \gamma  \left(4C^2-9 \tilde{Q}_c^2\right)}{C^2}}\bigg),\qquad \text{for} \quad \gamma  >1   
\end{split}
\end{equation}
and,

\begin{equation}\label{eq30}
\begin{split}
\tilde{q}_c^{2 \gamma }=\frac{3\times2^{\gamma-1 } \pi ^{-2 \gamma } C^{\gamma -1} S_c^{2 \gamma -1}}{(\gamma -1) (4 \gamma -1)}\bigg( S_c-\frac{2}{3}\pi  C\bigg), 
\end{split}
\end{equation}

\begin{equation}\label{eq31}
\begin{split}
\tilde{T}_c=\frac{2\pi  C (2 \gamma -1)+6 \gamma  S_c}{3 \pi ^{3/2} \sqrt{CS_c}  R (4 \gamma -1)}
\end{split}
\end{equation}

According to equation \eqref{eq29}, for $\gamma<1$ two distinct cases arise:
\begin{itemize}
    \item \textbf{Case I: $C>\frac{3}{2}\tilde{Q}_c$}
      
    In this regime, we obtain $S_c<\frac{2}{3 }\pi  C$. From equation \eqref{eq30}, the condition $\tilde{q}_c^2>0$ requires $\gamma>\tfrac{1}{4}$, while equation \eqref{eq31} demands $\gamma>\tfrac{1}{2}$ for $\tilde{T}_c>0$. Therefore, when $C>\frac{3}{2}\tilde{Q}_c$ and $\tfrac{1}{2}<\gamma<1$, critical points can exist for the black hole.

    \item \textbf{Case II: $C<\frac{3}{2}\tilde{Q}_c$}  
    
    In this case, equation \eqref{eq29} yields $S_c>\frac{2}{3 }\pi  C$. Equation \eqref{eq30} requires $\gamma<\tfrac{1}{4}$ for $\tilde{q}_c^2>0$, whereas equation \eqref{eq31} demands $\gamma>\tfrac{1}{2}$ for $\tilde{T}_c>0$. Since no common interval exists, there are no critical points for the black hole in this regime.
\end{itemize}
Furthermore, for $\gamma>1$, equation \eqref{eq29} gives $S_c<\frac{2}{3 }\pi  C$. However, equation \eqref{eq30} requires $\gamma<\tfrac{1}{4}$ for $\tilde{q}_c^2>0$, which is incompatible with $\gamma>1$. Consequently, no common interval exists, and critical points do not occur in this case either.\\

In addition, the heat capacity of the black hole can be derived from the following equation. This quantity plays a central role in characterizing the thermodynamic behavior of the system, as it directly reflects the regimes of stability and instability. A positive heat capacity corresponds to a thermodynamically stable phase, in which small perturbations in energy lead to controlled variations in temperature. Conversely, a negative heat capacity signals instability, indicating that the black hole cannot maintain equilibrium under thermal fluctuations. Thus, the evaluation of the heat capacity provides a fundamental criterion for distinguishing between different phases of black hole thermodynamics and for identifying possible phase transitions.
\begin{equation}\label{eq311}
\begin{split}
\tilde{\mathcal{C}}=\bigg(\frac{\partial E}{\partial \tilde{T}}\bigg)_{\tilde{Q}, \tilde{q}, C, \mathcal{V}}=\tilde{T} \bigg(\frac{\partial S}{\partial \tilde{T}}\bigg)_{\tilde{Q}, \tilde{q}, C, \mathcal{V}}
\end{split}
\end{equation}
By employing Equations \eqref{eq22} and \eqref{eq311}, the heat capacity of the black hole can be determined,
\begin{equation}\label{eq312}
\begin{split}
\tilde{\mathcal{C}}=\frac{2 S \left(\sqrt{\pi } \left(-2^{2 \gamma +1}\right) C^{\gamma } S^{2 \gamma } \left(\pi ^2 \tilde{Q}^2-S (4 \pi  C+3 S)\right)-2^{\gamma +2} \pi ^{2 \gamma +\frac{1}{2}} C S^2 \tilde{q}^{2 \gamma }\right)}{\sqrt{\pi } 2^{2 \gamma +1} C^{\gamma } S^{2 \gamma } \left(3 \pi ^2 \tilde{Q}^2+S (3 S-4 \pi  C)\right)+2^{\gamma +2} \pi ^{2 \gamma +\frac{1}{2}} (4 \gamma -1) C S^2 \tilde{q}^{2 \gamma }}
\end{split}
\end{equation}

\begin{figure}[h!]
 \begin{center}
 \subfigure[]{
 \includegraphics[height=5cm,width=6cm]{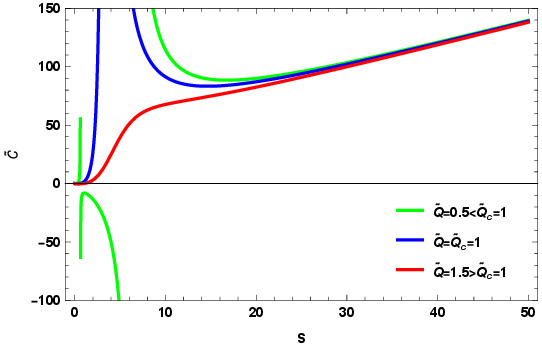}
 \label{figca}}
 \subfigure[]{
 \includegraphics[height=5cm,width=6cm]{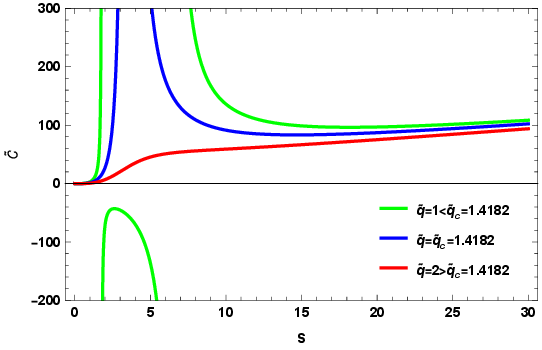}
 \label{figcb}}
  \subfigure[]{
 \includegraphics[height=5cm,width=6cm]{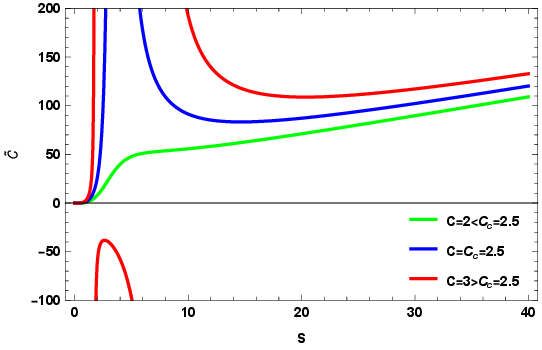}
 \label{figcc}}
 \caption{\small{ Heat Capacity $\tilde{\mathcal{C}}$ against $S$, with R=1 ,\quad $\gamma=0.6$, and \quad (a)  $\tilde{q}=1.41820$, $C=2.5$ \quad (b)  $\tilde{Q}=1$, $C=2.5$ \quad (c)  $\tilde{q}=1.41820$, $\tilde{Q}=1$  }}
 \label{figc}
\end{center}
 \end{figure}

From Fig. \ref{figc}, we observe that when $\tilde{Q}\geq \tilde{Q}_c$  , $\tilde{q}\geq \tilde{q}_c$ and $C \leq C_c$, the black hole remains in a thermodynamically stable state. In contrast, $\tilde{Q}< \tilde{Q}_c$  , $\tilde{q}< \tilde{q}_c$ and $C > C_c$, the system exhibits a negative heat capacity, signaling instability. On both sides of this unstable region, the entropy associated with the black hole’s heat capacity is positive, which suggests the presence of a first-order phase transition between small and large black hole configurations. This behavior motivates a more detailed investigation of the free energy landscape, which we present in the following section.

\section{ Exploring thermodynamic ensembles in the framework of CFT} \label{isec10}
Using the holographic dictionary developed in the preceding section, we investigate various thermodynamic ensembles---both canonical and grand canonical---within the CFT framework. In addition to the standard conjugate pair $(\tilde{T}, S)$, the CFT description of Einstein-Maxwell-Power-Yang-Mills-AdS Black Holes naturally involves several further pairs of thermodynamic variables: $(\tilde{\phi}, \tilde{Q})$, $(\tilde{\psi}, \tilde{q})$, $(p, \mathcal{V})$, and $(\mu, C)$.

Our analysis reveals the presence of phase transitions and critical phenomena in three distinct ensembles, namely at fixed $(\tilde{Q}, \tilde{q}, C, \mathcal{V})$, $(\tilde{\phi}, \tilde{q}, C, \mathcal{V})$, and $(\tilde{Q}, \tilde{\psi}, C, \mathcal{V})$. The corresponding free energies associated with these ensembles are denoted by $F$, $W$, and $G$, respectively,

\begin{equation}\label{eq32}
\begin{split}
& \text{fixed} \quad (\tilde{Q},\tilde{q} , C, \mathcal{V}):  \qquad  \tilde{F}\equiv E-\tilde{T}S=\tilde{\phi} \tilde{Q}+\frac{(3\gamma-1)}{2\gamma}\tilde{\psi} \tilde{q}+\mu C\\
&\text{fixed} \quad (\tilde{\phi},\tilde{q} ,C,\mathcal{V} ):  \qquad  \tilde{W}\equiv E-\tilde{T}S-\tilde{\phi} \tilde{Q}=\frac{(3\gamma-1)}{2\gamma}\tilde{\psi} \tilde{q}+\mu C   \\
%&\text{fixed} \quad (\tilde{Q},\tilde{q} , \mu, \mathcal{V}):  \qquad  \tilde{G}\equiv E-\tilde{T}S-\mu C=\tilde{\phi}\tilde{ Q}+\frac{(3\gamma-1)}{2\gamma}\tilde{\psi} \tilde{q}    
\end{split}
\end{equation}
In the following, we shall examine each case individually.

\subsection{ Ensemble at fixed $(\tilde{Q},\tilde{q} , C, \mathcal{V})$} 
In the canonical ensemble, fixing $\tilde{Q}$, $\tilde{q}$, $ C$, and $\mathcal{V}$ leads to the Helmholtz free energy as the relevant thermodynamic potential,
\begin{equation}\label{eq33}
\begin{split}
\tilde{F}=E-\tilde{T}S
\end{split}
\end{equation}
By the CFT first law \eqref{eq20}, the differential of  is given by,
\begin{equation}\label{eq34}
\begin{split}
d\tilde{F}= dE-Sd\tilde{T}-\tilde{T}dS=-Sd\tilde{T}+ \tilde{\phi}d\tilde{Q}+\tilde{\psi} d\tilde{q}+\mu dC-pd\mathcal{V}    
\end{split}
\end{equation}

By employing equations \eqref{eq17}, \eqref{eq22}, and \eqref{eq33}, we obtain,

\begin{equation}\label{eq35}
\begin{split}
\tilde{F}=\frac{2^{2-\gamma } \pi ^{2 \gamma } (1-4 \gamma ) C^{1-\gamma } \tilde{q}^{2 \gamma } S^{2-2 \gamma }+2 (3-4 \gamma ) \left(4 \pi  C S+3 \pi ^2 \tilde{Q}^2-S^2\right)}{16 \pi ^{3/2} (3-4 \gamma ) R \sqrt{C S}}
\end{split}
\end{equation}

 \begin{figure}[h!]
 \begin{center}
 \subfigure[]{
 \includegraphics[height=5cm,width=6cm]{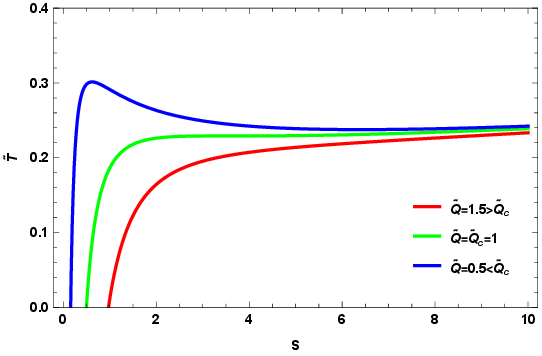}
 \label{fig1a}}
 \subfigure[]{
 \includegraphics[height=5cm,width=6cm]{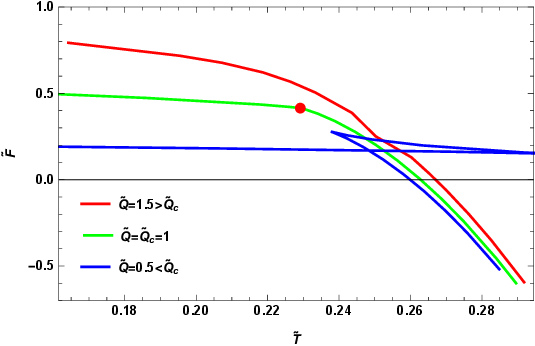}
 \label{fig1b}}
  \caption{\small{(a)  Temperature $\tilde{T}$ vs. entropy $S$   (b) Free energy $\tilde{F}$ vs. temperature $\tilde{T}$ plot  with R=1 ,\quad $\gamma=0.6$,\quad $\tilde{q}=1.41820$ \quad and, C=2.5 . The red dot marks the critical point.     }}
 \label{fig1}
\end{center}
 \end{figure}

\begin{figure}[h!]
 \begin{center}
 \subfigure[]{
 \includegraphics[height=5cm,width=6cm]{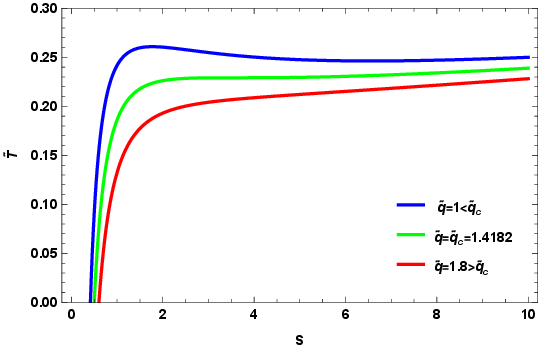}
 \label{fig2a}}
 \subfigure[]{
 \includegraphics[height=5cm,width=6cm]{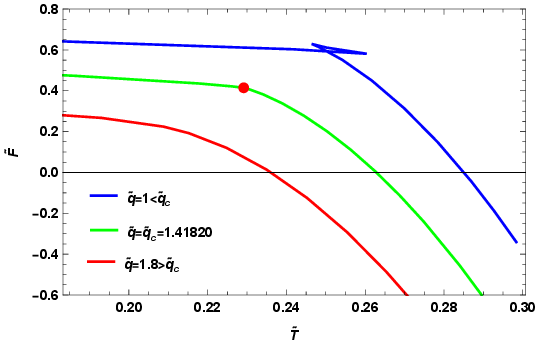}
 \label{fig2b}}
 \caption{\small{ (a)  Temperature $\tilde{T}$ vs. entropy $S$   (b) Free energy $\tilde{F}$ vs. temperature $\tilde{T}$ plot  with R=1 ,\quad $\gamma=0.6$ \quad $\tilde{Q}=1$  and, C=2.5 . The red dot marks the critical point.  }}
 \label{fig2}
\end{center}
 \end{figure}

 \begin{figure}[h!]
 \begin{center}
 \subfigure[]{
 \includegraphics[height=5cm,width=6cm]{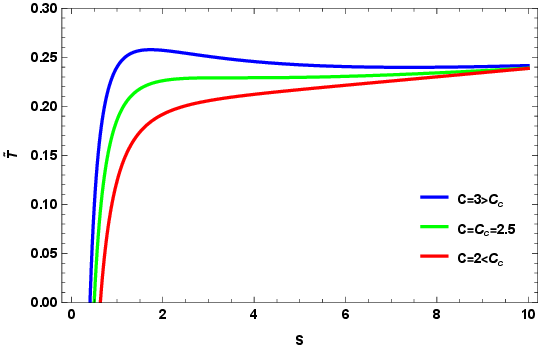}
 \label{fig3a}}
 \subfigure[]{
 \includegraphics[height=5cm,width=6cm]{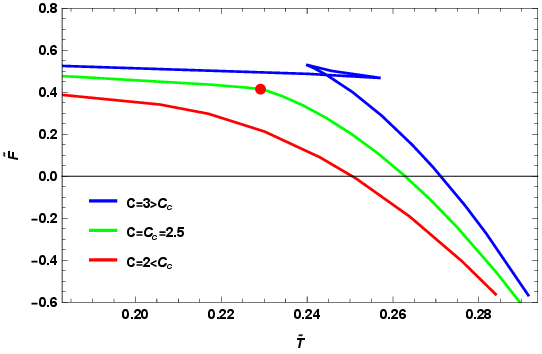}
 \label{fig3b}}
  \caption{\small{ (a)  Temperature $\tilde{T}$ vs. entropy $S$   (b) Free energy $\tilde{F}$ vs. temperature $\tilde{T}$ plot  with R=1 ,\quad $\gamma=0.6$\quad $\tilde{q}=1.41820$, and, $\tilde{Q}=1$ . The red dot marks the critical point.    }}
 \label{fig3}
\end{center}
 \end{figure}

%%%%%%%%%%%%%%%%%%%%%%%%%%%%%%%%%%%%%%%%%%%%%%%%%%%%
Using equations \eqref{eq22} and \eqref{eq35}, we construct the $\tilde{T}$–$S$ and $\tilde{F}$–$\tilde{T}$ diagrams to explore the thermodynamic properties of the Power–Maxwell–Young–Mills black hole within the holographic framework. These plots provide a systematic method for analyzing the phase structure of the system. In particular, by varying the key parameters $\tilde{q}$, $\tilde{Q}$, and $C$, we can identify signatures of both first- and second-order phase transitions, thereby shedding light on the thermodynamic behavior of this class of black holes.
For $\tilde{Q} < \tilde{Q}_{\text{c}}$ (blue curves Fig. \ref{fig1b}), the free energy develops a swallowtail structure, characteristic of a first-order phase transition between two stable thermodynamic branches. The nearly horizontal branch corresponds to a low-entropy phase, whereas the vertical branch represents a high-entropy phase. The intermediate segment connecting them exhibits negative heat capacity, rendering it unstable. At the critical point $\tilde{Q} = \tilde{Q}_{\text{c}}$ (green curve Fig. \ref{fig1b}), the system undergoes a continuous second-order transition. For $\tilde{Q} > \tilde{Q}_{\text{c}}$ (red curve Fig. \ref{fig1b}), the free energy remains smooth and analytic, with no phase transition present.
%%%%%%
Also, Fig. (\ref{fig2b})) reveals that for $\tilde{q} < \tilde{q}_{\text{c}}$, the free energy develops a swallowtail structure, signaling the presence of a first-order phase transition. At $\tilde{q} = \tilde{q}_{\text{c}}$, the system undergoes a continuous second-order transition. In contrast, when $\tilde{q} > \tilde{q}_{\text{c}}$, the free energy remains analytic and no phase transition occurs.
%%%%%%
Fig. \ref{fig3b} investigates the role of the central charge parameter $C$. For $C > C_c$, the system exhibits a clear first-order phase transition, while for $C < C_c$, no critical behavior is observed. The green curve, corresponding to $C = C_c$, marks the boundary between these regimes and identifies the second-order critical point.

Together, these diagrams demonstrate that the existence and nature of black hole phase transitions in this model are highly sensitive to the values of $\gamma$, $\tilde{q}$, $\tilde{Q}$, and $C$. The results provide a rich thermodynamic structure, including both first- and second-order transitions, governed by the interplay of gauge and gravitational parameters in the holographic setting.

 From Figures \ref{fig1a},\ref{fig2a}, and \ref{fig3a}, we observe that at sufficiently high entropies, variations in the parameters $\tilde{Q}$, $\tilde{q}$, and $C$ exert negligible influence on the system. Their corresponding temperature profiles coincide, leading to identical thermal behavior across these cases.

 \begin{figure}[h!]
 \begin{center}
 \subfigure[]{
 \includegraphics[height=5cm,width=6cm]{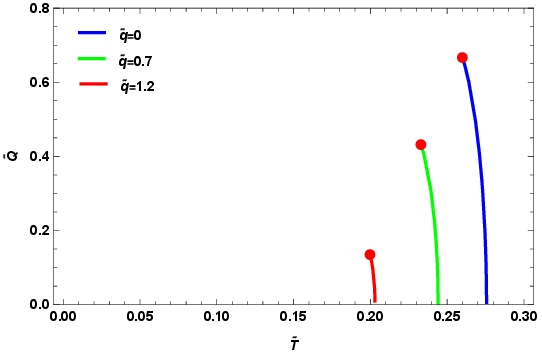}
 \label{figcox1a}}
 \subfigure[]{
 \includegraphics[height=5cm,width=6cm]{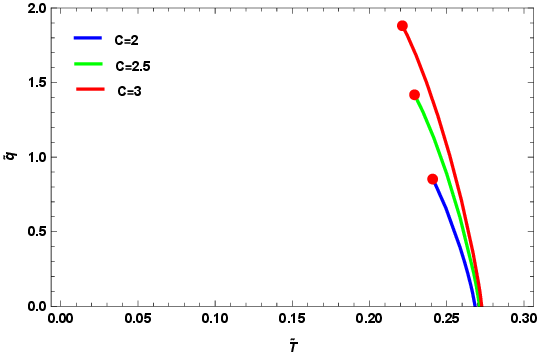}
 \label{figcox2b}}
  \subfigure[]{
 \includegraphics[height=5cm,width=6cm]{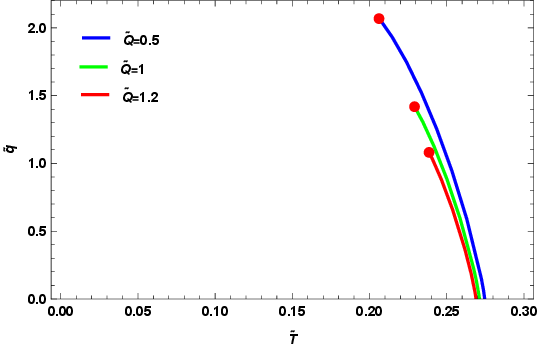}
 \label{figcox3c}}
 \caption{\small{ Coexistence lines  for the fixed $(\tilde{q},\tilde{Q}, \mathcal{V}, C)$ ensemble.  Low-entropy
and high-entropy coexistence curve for CFT thermal states on (a) $\tilde{Q}- \tilde{T}$ and (b),(c)$\tilde{q}-\tilde{T}$ phase diagram. The
parameters used here are (a) $R = 1, C=1, \gamma=0.6$, (b)  $R = 1,  \tilde{Q}=1, \gamma=0.6$, and (c)  $R = 1,  C=2.5, \gamma=0.6$.  For each value of (a) $q$ , (b) $C$, and (c) $\tilde{Q}$, the coexistence line represents a line of first-order phase transitions between low-entropy states (to the left of the line) and high-entropy states (to the right), and the line ends at a critical point where a second-order phase transition occurs at (a) $\tilde{Q}=\tilde{Q}_{c}$ and $\tilde{T}=\tilde{T}_{c}$ , (b), (c)  $\tilde{q}=\tilde{q}_{c}$ and $\tilde{T}=\tilde{T}_{c}$.   }}
 \label{figcox}
\end{center}
 \end{figure}
Figure \ref{figcox} displays the coexistence curves separating the low and high-entropy phases of the CFT on both the $\tilde{Q}-\tilde{T}$ and $\tilde{q}-\tilde{T}$ phase diagrams. These curves mark the boundary between the two phases, and crossing them corresponds to a first-order phase transition. In each diagram, the low-entropy phase lies to the left of the coexistence curve while the high-entropy phase lies to its right. Open circles denote the critical points; above these points the CFT no longer exhibits distinct phases.
In Figure \ref{figcox}, although $\tilde{Q}$ and $\tilde{q}$ in conformal field theory play the role of pressure to some extent, as in a van der Waals fluid, the coexistence lines in CFT have a negative slope, while in the P-T plane for a van der Waals fluid there is a positive slope.
%%%%%%%According to Figure \ref{figcox}, we find that the thermodynamic quantities $\tilde{q}$, $C$, and $\tilde{Q}$ affect the coexistence lines, such that in Figures \ref{figcox1a} and \ref{figcox3c}, as $\tilde{q}$ and $\tilde{Q}$ increases, the coexistence lines become smaller, while in Figure \ref{figcox2b}, as $C$ increases, the coexistence lines become larger.
Figure \ref{figcox} shows that the thermodynamic parameters \(\tilde{q}\), \(C\), and \(\tilde{Q}\) modify the phase-coexistence lines. In particular, in Figures \ref{figcox1a} and \ref{figcox3c} an increase in \(\tilde{q}\) or \(\tilde{Q}\) reduces the extent of the coexistence region (the coexistence curves contract), whereas in Figure \ref{figcox2b} an increase in \(C\) expands the coexistence region (the coexistence curves shift outward).
 
\subsection{ Ensemble at fixed $(\tilde{\phi},\tilde{q} ,C,\mathcal{V} )$} \label{isec3}
In the canonical ensemble, fixing $\tilde{\phi}$, $\tilde{q}$, $ C$, and $\mathcal{V}$ leads to the  free energy as the relevant thermodynamic potential,
\begin{equation}\label{eq37}
\begin{split}
\tilde{W}= E-\tilde{T}S-\tilde{\phi} \tilde{Q}
\end{split}
\end{equation}
By the CFT first law \eqref{eq20}, the differential of  is given by,
\begin{equation}\label{eq38}
\begin{split}
d\tilde{W}= dE-Sd\tilde{T}-\tilde{T}dS-\tilde{\phi}d\tilde{Q}-\tilde{Q}d\tilde{\phi}=-Sd\tilde{T}- \tilde{Q}d\tilde{\phi}+\tilde{\psi} d\tilde{q}+\mu dC-pd\mathcal{V}     
\end{split}
\end{equation}

By employing equations \eqref{eq17}, \eqref{eq22}, \eqref{eq23} and \eqref{eq37}, we obtain,

\begin{equation}\label{eq39}
\begin{split}
\tilde{W}=\frac{1}{2 R}\sqrt{\frac{C S}{\pi }}-\frac{S \left(\pi  C R^2 \tilde{\phi }^2+S\right)}{8 \pi ^{3/2} R \sqrt{C S}}+\frac{2^{-\gamma -2} \pi ^{2 \gamma -\frac{3}{2}} (4 \gamma -1) C^{\frac{1}{2}-\gamma } S^{\frac{3}{2}-2 \gamma } \tilde{q}^{2 \gamma }}{(4 \gamma -3) R}
\end{split}
\end{equation}

\begin{equation}\label{eq40}
\begin{split}
\tilde{T}=\frac{1}{2 R}\sqrt{\frac{C}{\pi  S}}+\frac{3 \sqrt{S}}{8 \pi ^{3/2} \sqrt{C} R}-\frac{2^{-\gamma -2} \pi ^{2 \gamma -\frac{3}{2}} C^{\frac{1}{2}-\gamma } S^{\frac{1}{2}-2 \gamma } \tilde{q}^{2 \gamma }}{R}-\frac{\sqrt{C} R \tilde{\phi }^2}{2 \sqrt{\pi  S}}
\end{split}
\end{equation}

\begin{figure}[h!]
 \begin{center}
 \subfigure[]{
 \includegraphics[height=5cm,width=6cm]{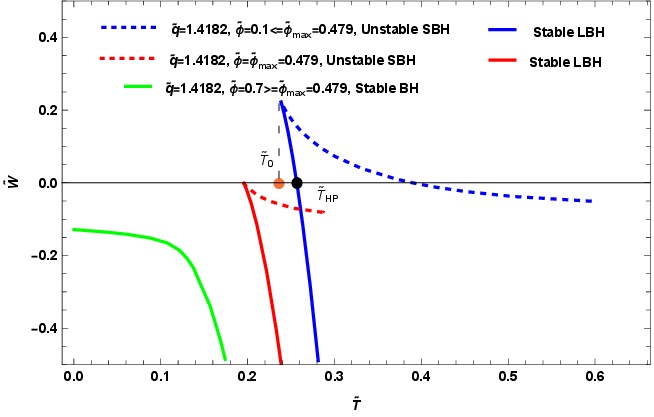}
 \label{figw1}}
 \subfigure[]{
 \includegraphics[height=5cm,width=6cm]{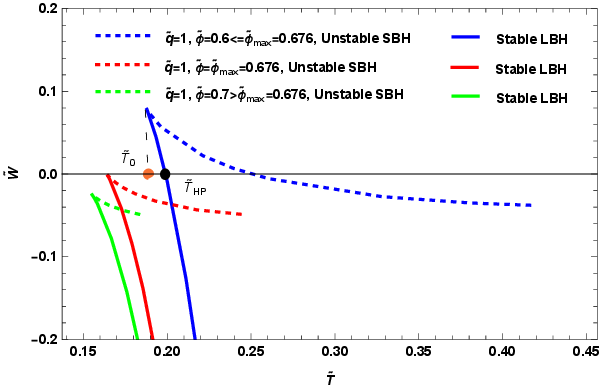}
 \label{figw2}}
  \caption{\small{$ \tilde{W} - \tilde{T}$ diagram of fixed $(\tilde{\phi},\tilde{q} ,C,\mathcal{V} )$ ensemble for $R=1, C=2.5, \gamma=0.6$. (a) $\tilde{q}=1.4182$ (b) $\tilde{q}=1 $.}}
 \label{figw0}
\end{center}
 \end{figure}

\begin{figure}[h!]
 \begin{center}
 \subfigure[]{
 \includegraphics[height=5cm,width=6cm]{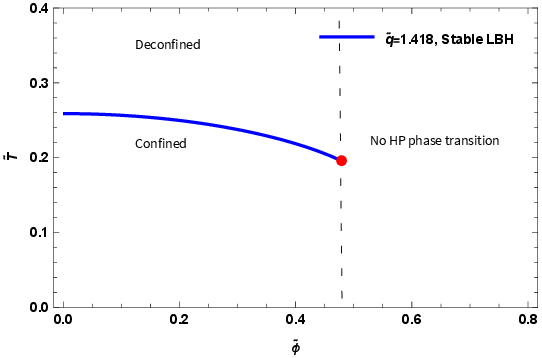}
 \label{figw3}}
 \subfigure[]{
 \includegraphics[height=5cm,width=6cm]{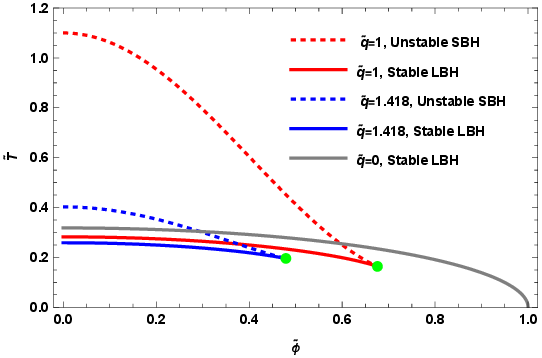}
 \label{figw4}}
  \caption{\small{  Co-existence diagram for $\tilde{T}-\tilde{\phi}$ with $R=1, C=2.5, \gamma=0.6$.  (a) $\tilde{q}=1.4182$ (b) $\tilde{q}=0,1,1.4182 $.  }}
 \label{figw00}
\end{center}
 \end{figure}

Figure \ref{figw0} illustrates the variation of free energy with temperature ($\tilde{W}(\tilde{T})$), as obtained from the formulations presented in Equations \eqref{eq22} and \eqref{eq39}.
In addition, when $\tilde{W}=0$, we present the phase diagram in the $\tilde{T}-\tilde{\phi}$ plane for the parameters $R=1, C=2.5, \gamma=0.6$. Figure  \ref{figw00} illustrates the coexistence curve, which serves as the boundary between distinct thermodynamic phases. This curve represents the line of phase transitions corresponding to confinement and deconfinement phenomena in CFT. 
 
As illustrated in Figure \ref{figw0}, the behavior of the free energy strongly depends on the value of $\tilde{\phi}$. For $\tilde{\phi} \geq \tilde{\phi}_{\text{max}}$, the free energy remains a single-valued function of the temperature, satisfying the condition $\tilde{W} \leq 0$, and the corresponding curve intersects the $\tilde{W}$-axis. This regime reflects a relatively simple thermodynamic structure, where the system evolves smoothly without exhibiting multiple competing phases. 
In contrast, when $\tilde{\phi} < \tilde{\phi}_{\text{max}}$, the free energy profile bifurcates into two distinct branches that merge at a cusp located at $T_0$.
The upper branch corresponds to small black hole (SBH) states, which are characterized by low entropy and negative specific heat. Such states are thermodynamically unstable. The lower branch, on the other hand, represents large black hole (LBH) states with high entropy and positive specific heat. These configurations are thermodynamically stable.
From the plot displayed in Figure \ref{figw0}, it is evident that the free energy of the lower branch changes sign exactly at $\tilde{W }= 0$. This behavior indicates the presence of a first-order phase transition. For $\tilde{W} < 0$, the ensemble is dominated by the high-entropy ``deconfined'' phase, whereas for $\tilde{W} > 0$ the thermodynamically stable configuration is the ``confined'' phase. The transition between these regimes corresponds to a generalized Hawking-Page phenomenon, which relates large AdS black holes to the AdS background filled with thermal radiation \cite{ref4,ref5}. 
Based on the Fig. \ref{figw3}, we observe that the phase transition occurs only for  $\tilde{\phi} < \tilde{\phi}_{\max}$, while for $\tilde{\phi} > \tilde{\phi}_{\max}$ 
no Hawking-Page transition takes place and the deconfined large black hole phase  remains dominant. In the Fig. \ref{figw4}, the influence of the Yang-Mills charge parameter 
$\tilde{q}$ becomes evident: as $\tilde{q}$ increases, the range of coexistence shrinks, and the domain in which the first-order phase transition is realized 
becomes narrower. This indicates that higher Yang-Mills charge suppresses the  confinement/deconfinement transition in the holographic setup.

The temperature exhibits its minimum with respect to $S$ precisely at the cusp,  i.e.
\begin{equation}\label{eq40}
\begin{split}
\bigg(\frac{\partial \tilde{T}}{\partial S}\bigg)_{\tilde{\phi}}=0 \quad \rightarrow \quad S_0=S_{cusp}  \quad \rightarrow \quad  \tilde{T}_0=\tilde{T}_{cusp}
\end{split}
\end{equation}
Based on Equations \eqref{eq22}, \eqref{eq39} and \eqref{eq40}, neither the minimum temperature ($T_0$) nor the Hawking–Page transition temperature ($\tilde{W}=0 \rightarrow \tilde{T}_{HP}$) can be determined in closed analytic form. Consequently, these quantities must be evaluated numerically for various values of, as summarized in the table below.
%q=1.4182

\begin{table}[h!]
    \centering
    \setlength{\arrayrulewidth}{0.5pt}
    \setlength{\tabcolsep}{10pt}
    \renewcommand{\arraystretch}{1.3}
    \begin{tabular}{@{}c c c c c@{}}
        \toprule
        \rowcolor{headerblue}
        \textcolor{white}{$\tilde{q}$} &
        \textcolor{white}{$\tilde{\phi}$} &
        \textcolor{white}{$\tilde{T}_0$} &
        \textcolor{white}{$\tilde{T}_{\text{HP}}$ (LBH)} &
        \textcolor{white}{$\Delta \tilde{T} = \tilde{T}_{\text{HP}} - \tilde{T}_0$} \\
        \midrule
        1.4182 & 0.1 & 0.2379 & 0.2566 & 0.0187 \\
        \rowcolor{rowgray}
        1.4182 & 0.2 & 0.2327 & 0.2497 & 0.0170 \\
        1.4182 & 0.3 & 0.2236 & 0.2376 & 0.0140 \\
        \rowcolor{rowgray}
        1.4182 & 0.4 & 0.2101 & 0.2187 & 0.0086 \\
        1.4182 & 0.479 & 0.1956 & 0.1956 & 0.0000 \\
        \bottomrule
    \end{tabular}
    \caption{
        Numerical evaluation of the minimum temperature $\tilde{T}_{0}$
        and the Hawking-Page transition temperature $\tilde{T}_{\text{HP}}$
        for various values of $\tilde{\phi}$. Fixed parameters:
        $R=1$, $C=2.5$, $\gamma=0.6$, and $\tilde{q}=1.4182$.}
    \label{P3}
\end{table}

%q=1

\begin{table}[h!]
    \centering
    \setlength{\arrayrulewidth}{0.5pt}
    \setlength{\tabcolsep}{12pt}
    \renewcommand{\arraystretch}{1.3}
    \begin{tabular}{@{}c c c c c@{}}
        \toprule
        \rowcolor{headerblue}
        \textcolor{white}{$\tilde{q}$} &
        \textcolor{white}{$\tilde{\phi}$} &
        \textcolor{white}{$\tilde{T}_0$} &
        \textcolor{white}{$\tilde{T}_{\text{HP}}$ (LBH)} &
        \textcolor{white}{$\Delta\tilde{T} = \tilde{T}_{\text{HP}} - \tilde{T}_0$} \\
        \midrule
        1 & 0.1 & 0.2516 & 0.2805 & 0.0289 \\
        \rowcolor{rowgray}
        1 & 0.2 & 0.2469 & 0.2747 & 0.0278 \\
        1 & 0.3 & 0.2387 & 0.2647 & 0.0260 \\
        \rowcolor{rowgray}
        1 & 0.4 & 0.2267 & 0.2498 & 0.0231 \\
        1 & 0.675 & 0.1646 & 0.1646 & 0.0000 \\
        \bottomrule
    \end{tabular}
    \caption{
        Numerical evaluation of the minimum temperature $\tilde{T}_{0}$
        and the Hawking-Page transition temperature $\tilde{T}_{\text{HP}}$
        for different values of $\tilde{\phi}$, obtained with the fixed 
        parameters $R=1$, $C=2.5$, $\gamma=0.6$, and $\tilde{q}=1$.}
    \label{P4}
\end{table}

According to Tables \ref{P3} and \ref{P4}, the Yang–Mills charge parameter $(\tilde{q})$ significantly influences both the Hawking–Page transition temperature and the minimum temperature. As $\tilde{q}$ increases while keeping $\tilde{\phi}$ constant, the values of these two temperatures ($\tilde{T}_{HP}$ and $\tilde{T}_{0}$) decrease.
Furthermore, as $\tilde{q}$ increases, the value of $\Delta \tilde{T}=\tilde{T}_{HP}-\tilde{T}_0$ decreases, indicating that the “confined” phase is becoming smaller.

\newpage
\section{Conclusion} \label{sec5}
In this research, we employed the framework of extended thermodynamics and holographic duality to investigate the phase properties and stability of Einstein-Maxwell-power-Yang-Mills black holes in Anti-de Sitter spacetime from the perspective of their dual boundary CFT. By introducing the central charge \(C\) as a fundamental thermodynamic variable in the first law (Eq. \eqref{eq20}), and defining several thermodynamic potentials corresponding to different ensembles (Eq. \eqref{eq32}), we demonstrated that the phase landscape of this system is governed by a complex interplay of theoretical parameters: the electric charge \(\tilde{Q}\), the Yang-Mills charge \(\tilde{q}\), the electric potential \(\tilde{\phi}\), and the central charge \(C\). In the following, we provide a detailed comparative analysis and discussion of the key results.
In this ensemble (Canonical Ensemble (Fixed \(\tilde{Q}\), \(\tilde{q}\), \(C\), \(\mathcal{V}\)): Helmholtz Free Energy \(\tilde{F}\)), the system exhibits behavior remarkably similar to a van der Waals fluid. The temperature \(\tilde{T}\) as a function of entropy \(S\) (Figs. \ref{fig1a}, \ref{fig2a}, \ref{fig3a}) shows the characteristic S-shaped curve, indicative of instability and phase transition. This behavior stems directly from the temperature formula given in Eq. \eqref{eq22}. The condition for the existence of critical points (\(\partial\tilde{T}/\partial S=0\) and \(\partial^2\tilde{T}/\partial S^2=0\)) leads to the complex relations for critical entropy \(S_c\) (Eq. \eqref{eq29}) and critical charge \(\tilde{q}_c\) (Eq. \eqref{eq30}).
Figs. \ref{fig1b}, \ref{fig2b}, and \ref{fig3b} clearly show that a first-order phase transition (swallowtail structure) occurs only within specific parameter ranges: For \(\tilde{Q} < \tilde{Q}_c\) (Fig. \ref{fig1b}). For \(\tilde{q} < \tilde{q}_c\) (Fig. \ref{fig2b}).For \(C > C_c\) (Fig. \ref{fig3b}).
At the critical values (\(\tilde{Q} = \tilde{Q}_c\), \(\tilde{q} = \tilde{q}_c\), \(C = C_c\)), the swallowtail structure vanishes and the system undergoes a continuous second-order phase transition. This behavior indicates that the charges \(\tilde{Q}\) and \(\tilde{q}\) play the role of generalized pressures, while the central charge \(C\) acts like an inverse generalized volume in analogy with the van der Waals system. The coexistence curves in the \(\tilde{Q}-\tilde{T}\) and \(\tilde{q}-\tilde{T}\) planes (Fig. \ref{figcox}) demarcate the boundary between the low-entropy (small black hole) and high-entropy (large black hole) phases. A particularly intriguing feature is the negative slope of these curves (unlike the positive slope in the P-T diagram of a van der Waals fluid). This significant difference highlights the distinct nature of "pressure" in the boundary field theory; here, increasing \(\tilde{Q}\) or \(\tilde{q}\) (analogous to increasing pressure) shrinks the coexistence region and suppresses the more condensed (low-entropy) phase (Figs. \ref{figcox1a} and \ref{figcox3c}). Conversely, increasing \(C\) (associated with a decreasing length scale \(\ell\)) expands the coexistence region (Fig. \ref{figcox2b}).\\\\
 
This ensemble (Mixed or Semi-Grand Canonical Ensemble (Fixed \(\tilde{\phi}\), \(\tilde{q}\), \(C\), \(\mathcal{V}\)): Potential \(\tilde{W}\)), where the electric potential \(\tilde{\phi}\) is held fixed, reveals a fundamentally different physics. Here, the primary phase transition is a generalized Hawking-Page transition between a "confined" and a "deconfined" phase.
The \(\tilde{W}\) vs. \(\tilde{T}\) plot (Fig. \ref{figw0}) shows a bifurcating structure with a cusp at the minimum temperature \(\tilde{T}_0\). The upper (unstable) branch corresponds to small black holes (SBHs), and the lower (stable) branch corresponds to large black holes (LBHs).
The first-order phase transition point is where the free energy of the stable (LBH) branch becomes zero: \(\tilde{W} = 0 \rightarrow \tilde{T}_{HP}\). For \(\tilde{T} < \tilde{T}_{HP}\), the stable phase is the "confined" state (likely the AdS thermal radiation background) with \(\tilde{W} > 0\). For \(\tilde{T} > \tilde{T}_{HP}\), the stable phase is the "deconfined" state (large AdS black hole) with \(\tilde{W} < 0\).
The Hawking-Page transition only occurs for \(\tilde{\phi} < \tilde{\phi}_{\text{max}}\) (Fig. \ref{figw3}). For \(\tilde{\phi} > \tilde{\phi}_{\text{max}}\), \(\tilde{W}\) is always negative, and the deconfined (LBH) phase dominates at all temperatures. This implies that a high electric potential can completely suppress the confined phase.
The numerical data presented in Tables \ref{P3} and \ref{P4} confirm a key conclusion: Increasing the Yang-Mills charge \(\tilde{q}\) weakens the Hawking-Page transition. Keeping \(\tilde{\phi}\) constant and increasing \(\tilde{q}\): Both temperatures \(\tilde{T}_0\) and \(\tilde{T}_{HP}\) decrease.
The temperature gap \(\Delta\tilde{T} = \tilde{T}_{HP} - \tilde{T}_0\) narrows (from 0.0289 for \(\tilde{q}=1\) to 0.0187 for \(\tilde{q}=1.4182\) at \(\tilde{\phi}=0.1\)). The critical value \(\tilde{\phi}_{\text{max}}\) (where \(\Delta\tilde{T}\) vanishes) decreases (from 0.675 for \(\tilde{q}=1\) to 0.479 for \(\tilde{q}=1.4182\)).
These observations indicate that the presence of the non-linear Yang-Mills field (\(\gamma \neq 1\)) not only acts as an additional charge source but also qualitatively alters the stability of the AdS background relative to the large black hole, narrowing the window for transition between confined and deconfined phases.\\
The results show that for a given set of parameters, the most stable configuration can differ depending on the thermodynamic ensemble used to describe the system. In the canonical ensemble, the competition is between SBH and LBH. In the mixed ensemble, the competition is between the AdS background (confined) and the LBH (deconfined). Differentiation of the internal energy \(E\) (Eq. \eqref{eq17}) led to the definition of the chemical potential \(\mu\) for the central charge \(C\) (Eq. \eqref{eq21}) and the boundary pressure \(p\) (Eq. \eqref{eq26}). These quantities describe the dynamics of the boundary field theory in its own physical space (with volume \(\mathcal{V}=4\pi R^2\)), enabling direct matching with CFT thermodynamics.
The heat capacity formula \(\tilde{\mathcal{C}}\) (Eq. \eqref{eq312}) and its plots (Fig. \ref{figc}) directly confirm the regions of instability (\(\tilde{\mathcal{C}}<0\)) that appeared as intermediate branches in the free energy plots. The perfect agreement between the local (\(\tilde{\mathcal{C}}\)) and global (free energy) analyses validates the consistency of the calculations.\\

These results open a new avenue for exploring the interplay between gravitational thermodynamics, non-Abelian gauge field theories, and phase phenomena in conformal field theories.

\end{document}